\documentclass[pdflatex,sn-nature]{sn-jnl}

\usepackage{graphicx}
\usepackage{multirow}
\usepackage{amsmath,amssymb,amsfonts}
\usepackage{amsthm}
\usepackage{mathrsfs}
\usepackage[title]{appendix}
\usepackage{xcolor}
\usepackage{textcomp}
\usepackage{manyfoot}
\usepackage{booktabs}
\usepackage{algorithm}
\usepackage{algorithmicx}
\usepackage{algpseudocode}
\usepackage{listings}
\usepackage{lmodern}
\usepackage{anyfontsize}
\usepackage{lipsum}
\usepackage{pdfpages}
\usepackage{booktabs}
\usepackage{makecell}
\usepackage{siunitx}
\usepackage{tabularx,booktabs}
\usepackage{makecell}
\usepackage{parskip}

\DeclareSIUnit{\milipascal}{mPa}

\begin{document}

\title[Article Title]{Amorphous Silicon-Carbide Photonics for Ultrasound Imaging}

\author*[1]{\fnm{R. Tufan} \sur{Erdogan}}\email{r.t.erdogan@tudelft.nl}

\author[2]{\fnm{Bruno} \sur{Lopez-Rodriguez}}\email{b.lopezrodriguez@tudelft.nl}

\author[1]{\fnm{Wouter J.} \sur{Westerveld}}\email{w.j.westerveld@tudelft.nl}

\author[1]{\fnm{Sophinese} \sur{Iskander-Rizk}}\email{s.iskander-rizk@tudelft.nl}

\author[1]{\fnm{Gerard J.} \sur{Verbiest}}\email{g.j.verbiest@tudelft.nl}

\author[2]{\fnm{Iman} \sur{Esmaeil Zadeh}}\email{i.esmaeilzadeh@tudelft.nl}

\author*[1]{\fnm{Peter G.} \sur{Steeneken}}\email{p.g.steeneken@tudelft.nl}

\affil[1]{\orgdiv{Department of Precision and Microsystems Engineering}, \orgname{Faculty of Mechanical Engineering,Delft University of Technology}, \orgaddress{\street{Mekelweg 2}, \city{Delft}, \postcode{2628 CD}, \state{}, \country{The Netherlands}}}

\affil[2]{\orgdiv{Department of Imaging Physics (ImPhys)}, \orgname{Faculty of Applied Sciences, Delft University of Technology}, \orgaddress{\street{Lorentzweg 1}, \city{Delft}, \postcode{2628 CJ}, \state{}, \country{The Netherlands}}}

\abstract{Photonic ultrasound sensors offer unparalleled potential for improving spatial and temporal resolution in ultrasound imaging applications due to their size-independent noise figure, high sensitivity, and broad bandwidth characteristics. The exploration of optical materials in ultrasound sensing has attracted significant interest due to their potential to further enhance device performance, fabrication compatibility, and long-term stability. This work introduces amorphous silicon carbide (a-SiC) as a material for optical ultrasound sensing. a-SiC offers a balanced combination of optical confinement, low propagation loss, and high material stability, enabling sensitive and miniaturized microring sensors. We demonstrate a compact ultrasound detection system with a linear array of 20 transducers coupled to a single bus waveguide. The sensors feature a maximum optical finesse of 1320 and an intrinsic platform sensitivity of \SI{78}{\femto\metre\per\kilo\pascal}, leading to a noise equivalent pressure density below \SI{55}{\milipascal\per\sqrt\hertz}, which is calibrated to a hydrophone from \SI{3.36}{\mega\hertz} to \SI{30}{\mega\hertz}. Additionally, we perform high-resolution ultrasound imaging of fine structures, highlighting the viability of these sensors in real-world applications. Moreover, its low deposition temperature allows a-SiC material for optical ultrasound sensors to be integrated on most substrates. These results position a-SiC as a promising material for optical ultrasound sensing, offering a miniaturized, low-loss, and high-sensitivity solution for next-generation photoacoustic imaging applications.}

\maketitle
\newpage
\section*{Introduction}\label{sec1}

Medical ultrasound imaging is an indispensable tool for disease diagnosis. It enables capturing structural images of tissue deep under the skin non-invasively. For this purpose, pulses of ultrasound waves are sent deep into the tissue, and the echoes generated by reflections due to variations in acoustic impedance are recorded at the skin to create an image. However, because different tissues have almost the same acoustic impedance, these images contain little information on the variations in the type of tissue. Therefore, photoacoustic imaging (PAI) has been introduced, which gathers valuable information on the molecular composition of tissue by photoselective generation of ultrasound waves, making it a promising candidate to become a mainstream imaging modality in clinical settings \cite{attia_review_2019,steinberg_photoacoustic_2019}. 

Reconstructing 2D or 3D acoustic or photoacoustic tomography (PAT) images requires recording the ultrasound wavefield at multiple locations \cite{zhou_tutorial_2016}. This can be achieved either by scanning a single sensor across the area of interest or by using an array of sensors distributed over a plane. Current PAT imaging systems typically employ piezoelectric ultrasound sensor arrays or matrices to facilitate image reconstruction without the need for scanning \cite{noauthor_ithera_nodate}. However, piezoelectric detectors face inherent limitations in sensitivity, bandwidth, and scalability \cite{dong_opt}. Their sensitivity decreases as their size is reduced, their bandwidth is limited by the piezoelectric response bandwidth, and each individual piezoelectric element requires a dedicated cable, which constrains integration density and application potential. To our knowledge, the smallest reported pitch for a piezoelectric detector array is \qty{100}{\micro\metre} \cite{guo_piezo_pitch}. Since the spatial resolution in PAT is directly influenced by the detectors' element size, kerf, and bandwidth, which limits the spatial frequencies allowed in the grid, this restriction in piezoelectric sensors ultimately constrains the details in reconstructed images.

Although commercial piezoelectric hydrophones can detect high-frequency ultrasound signals and enable scanning ultrasound tomography with small scanning step sizes in a time-consuming fashion, they exhibit trade-offs in sensitivity when miniaturized. For example, commercial piezoelectric needle hydrophones with diameters of \qty{0.04}{\mm} and \qty{0.2}{\mm} have noise equivalent pressures (NEP) of \qty{10}{\kilo\pascal} and \qty{1.1}{\kilo\pascal} with operating bandwidths of \qty{60}{\mega\hertz} and \qty{40}{\mega\hertz} \cite{pa40um,pa200um}, respectively. In contrast, optical ultrasound detectors have demonstrated NEP values more than ten times lower \cite{shnaiderman_submicrometre_2020, westerveld_sensitive_2021} and operational bandwidths more than three times higher \cite{shnaiderman_submicrometre_2020, hazan_silicon-photonics_2022} compared to aforementioned hydrophones while maintaining sensor sizes smaller than \qty{20}{\micro\metre}.

Accordingly, optical ultrasound detection using integrated photonic circuits is a promising technique with the potential to meet the PAI requirements in miniaturization, scalability, sensitivity, and bandwidth \cite{rosenthal_embedded_2014}. These sensors detect acoustic pressure via the effect on the refractive index and the deformation of the optical structure due to acoustic pressure \cite{tsesses_modeling_2017}. To maximize detection performance while keeping the sensor size miniature, the optical response to acoustic pressure, such as resonance shift per impinging pressure, should be as high as possible. In addition, low optical propagation loss and high finesse are critical parameters that directly influence the sensor’s sensitivity, image resolution, and parallelization of the readout. Several optical waveguide geometries and materials were demonstrated using single miniature point-like sensors for integrated photonic ultrasound sensing. Polymer \cite{ultrabroadban_song}, silicon \cite{westerveld_optical_2019}, and silicon-nitride \cite{nagli_silicon_2023} platforms have been employed to create an optical ring resonator and have proven the viability of integrated photonic ultrasound detection for photoacoustic imaging applications. Among these demonstrations, silicon waveguides with polymer top cladding are promising since their integration with the CMOS-fabricated electronics merely requires a small additional step of spin-coating. To this date, polydimethylsiloxane (PDMS), as a top cladding polymer, has been found to be a highly photoelastic material \cite{BaoReview} under ultrasound pressure and is used in ultrasound sensing demonstrations\cite{hazan_silicon-photonics_2022, nagli_silicon_2023,erdogan_PDMS}.  Although single sensors could satisfy some of the needs in photoacoustic microscopy applications, a balanced demonstration of sensor arrays that address all requirements at once while maintaining small size, high sensitivity, and broad bandwidth, with a scalable, low-cost fabrication architecture, is still an open goal with high potential impact in photoacoustic tomography applications, hence the need for exploration of new waveguide materials is required.
 
\begin{figure}[!htb]
\centering
\includegraphics[width=1\textwidth]{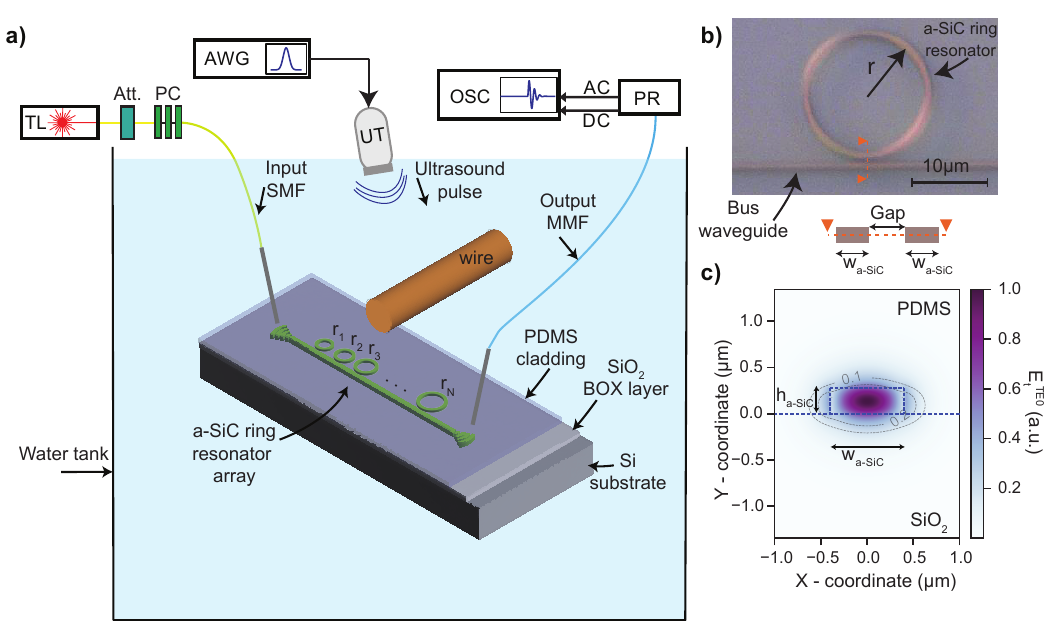}
\caption{Ultrasound sensing with an array of a-SiC ring resonators. \textbf{a)} Schematic of the integrated photonic sensor array chip and measurement setup. TL: tunable laser, Att: optical attenuator, PC: polarization control paddles, SMF: single-mode fiber, MMF: multi-mode fiber, AWG: arbitrary waveform generator, UT: ultrasound transducer, OSC: oscilloscope, PR: photoreceiver, AC, and DC: alternating and direct current, respectively.    \textbf{b)} Microscope image of a single PDMS-coated amorphous SiC micro ring resonator with $r =$ \qty{8}{\micro\metre}, $w_{a-SiC} =$  \qty{0.8}{\micro\metre},  and $gap =$ \qty{0.45}{\micro\metre}. The cross-sectional schematic of the bus waveguide and ring waveguide is given at the bottom of the microscope image, and parameters associated with the cross-section are indicated.  \textbf{c)} Calculated total electric field amplitude of TE0-mode for the cross section with an a-SiC core height of \qty{280}{\nano\metre} and a width of \qty{800}{\nano\metre}. The material boundaries are indicated with blue dashed lines. The cover and bottom cladding materials are PDMS and SiO\textsubscript{2}, respectively. The contour lines indicate the amplitude of the electric field, showing that the electric field is sensitive to variations in the cladding refractive index, primarily within approximately \qty{500}{\nano\metre} of the waveguide core, with the sensitivity gradually diminishing beyond this region.}
	\label{figConcept}
\end{figure}

Recently, amorphous silicon carbide (a-SiC) has been introduced as a new material for realizing integrated photonic ring resonators. It was demonstrated that a-SiC ring resonators possess high intrinsic-quality-factors, reaching $Q=5.7\times10^5$ with propagation losses as low as \qty{0.78}{\decibel\per\centi\metre} \cite{bruno_acs}. The a-SiC layer offers a refractive index $n=2.55$ at \qty{1550}{\nano\metre} wavelength, which can be adjusted by tuning its composition. The refractive index of a-SiC is higher than that of polymer, silicon nitride, and chalcogenide-based waveguides, but lower than that of silicon-on-insulator waveguides. The intermediate value of the refractive index of a-SiC offers a good compromise between size (higher $n$ enables smaller ring diameter) and propagation losses (smaller $n$ results in lower propagation losses).

Another advantage of a-SiC films is that they can be grown with high optical quality at temperatures as low as \qty{150}{\degreeCelsius}. This allows a-SiC to be integrated in the backend-of-line of standard complementary metal-oxide-semiconductor (CMOS) processes or on flexible material platforms that cannot withstand high temperatures \cite{geiger2020flexible, huang2024flexible}. This offers two prospects; firstly, integrating thousands of sensitive photonic ultrasound sensors with opto-electronic detection, acoustic signal processing, and imaging electronics in a single microchip can be achieved through CMOS integration. Secondly, the fabrication of the a-SiC waveguides can be performed on non-silicon substrates. For example, flexible substrates can be used to match the acoustic impedance of tissue or to integrate the backing layer in ultrasound applications. Additionally, transparent substrates, such as glass, can be used to allow photoacoustic excitation by light that passes through the sensor’s substrate \cite{li_transparent_2014}. Furthermore, the nonlinear optical Kerr coefficient of a-SiC is ten times higher than that of silicon nitride and crystalline silicon carbide \cite{ye2019low, lu2014optical}. This nonlinear property allows generation of on-chip frequency combs \cite{wang2021high}, which could enable on-chip integrated parallel interrogation methods \cite{pan_parallel_2023} and can increase the speed and imaging quality of these sensors while reducing their size.

Here, we demonstrate the potential of integrated optical ultrasound sensors using a-SiC ring resonators and resonator arrays by showing their performance for ultrasound tomography. For this purpose, we use a linear array of 20 a-SiC ring resonators, with a diameter of \qty{16}{\micro\metre} at a pitch of \qty{30}{\micro\metre}, that is coupled to a single bus waveguide. The a-SiC ring resonators feature a low noise equivalent pressure density (NEPD) below \SI{55}{\milli\pascal\per\sqrt\hertz} at a noise equivalent pressure (NEP) of \SI{345}{\pascal}, over a bandwidth of \qty{26.6}{\mega\hertz}. Despite their small dimension (sensor radius of \qty{8}{\micro\metre}), the ring resonators have a high intrinsic waveguide sensitivity  (\qty{78}{\femto\metre\per\kilo\pascal}) and Q-factor ($>1\times10^5$). To demonstrate its potential for ultrasound tomography, we use the linear array to capture 2D cross-sectional ultrasound images of a \qty{50}{\micro\metre} diameter aluminum wire and two  \qty{125}{\micro\metre} diameter glass fibers, without scanning the position of the sensor array. The image quality provides confidence that the a-SiC platform is suitable for sensitive and broadband ultrasound imaging, in advanced applications like photoacoustic tomography.

\section*{Results}\label{sec2}

Figure \ref{figConcept}a depicts the ultrasound characterization and imaging setup. The system enables ultrasound imaging with a-SiC circular microring resonator arrays, coupled to a straight bus waveguide. The light is coupled into and out of the chip using two focusing grating couplers, allowing access to the passive photonic circuitry through fibers positioned on both sides of the chip. The measurement setup includes a tunable laser to interrogate the sensors across different wavelengths. A photoreceiver connected to the output fiber converts the optical signals to electrical signals. The analog output from the photoreceiver is digitized using an oscilloscope triggered by the same arbitrary waveform generator that generates the transmitted ultrasound signals.

A microscope image of the polydimethylsiloxane cladded a-SiC ring resonator is provided in Fig. \ref{figConcept}b. The resonator comprises a circular ring waveguide with a radius, $r$, and a straight bus waveguide placed at a gap distance from the circular ring resonator. The parameters defining the geometry of the single-ring resonator are provided in Figs. \ref{figConcept}b and c. To evaluate and compare the optical performance and ultrasound sensitivity, we fabricated resonators with a waveguide width of \qty{800}{\nano\metre} and a coupling gap ranging from \qty{300}{\nano\metre} to \qty{650}{\nano\metre} for ring radii of \qty{6}{\micro\metre} and \qty{8}{\micro\metre} at waveguide height \qty{280}{\nano\metre}. Waveguide widths and heights of the resonator and the bus waveguides are kept the same throughout the study. 

The main physical mechanism by which ultrasound can be sensed with integrated optical waveguides is the dependence of the refractive index of the waveguide materials on the incoming pressure. In the case of a highly photo-elastic polymer cladding, such as PDMS, the changes in the refractive index of the polymer cladding dominate the change in the overall effective index due to incoming pressure \cite{hazan_silicon-photonics_2022,erdogan_PDMS}. To assess this effective refractive index, in Fig. \ref{figConcept}c, we provide a FEM simulation of the electric field of the TE0 mode and a schematic of the cross-section of the a-SiC waveguide used in this study. It comprises an a-SiC layer in the core, PDMS, and silicon dioxide in the top and bottom cladding layers, respectively. The simulation shows that the electric field distribution extends beyond the waveguide core boundary, retaining approximately 10\% of its amplitude at a maximum distance of about $\pm$\qty{500}{\nano\metre} from the center of the waveguide core for the plotted geometry. This evanescent field makes the waveguide sensitive to ultrasound pressure-induced changes in the refractive index of the PDMS via the photoelasticity of PDMS.

\begin{figure}[!htb]
\centering
\includegraphics[width=1\textwidth]{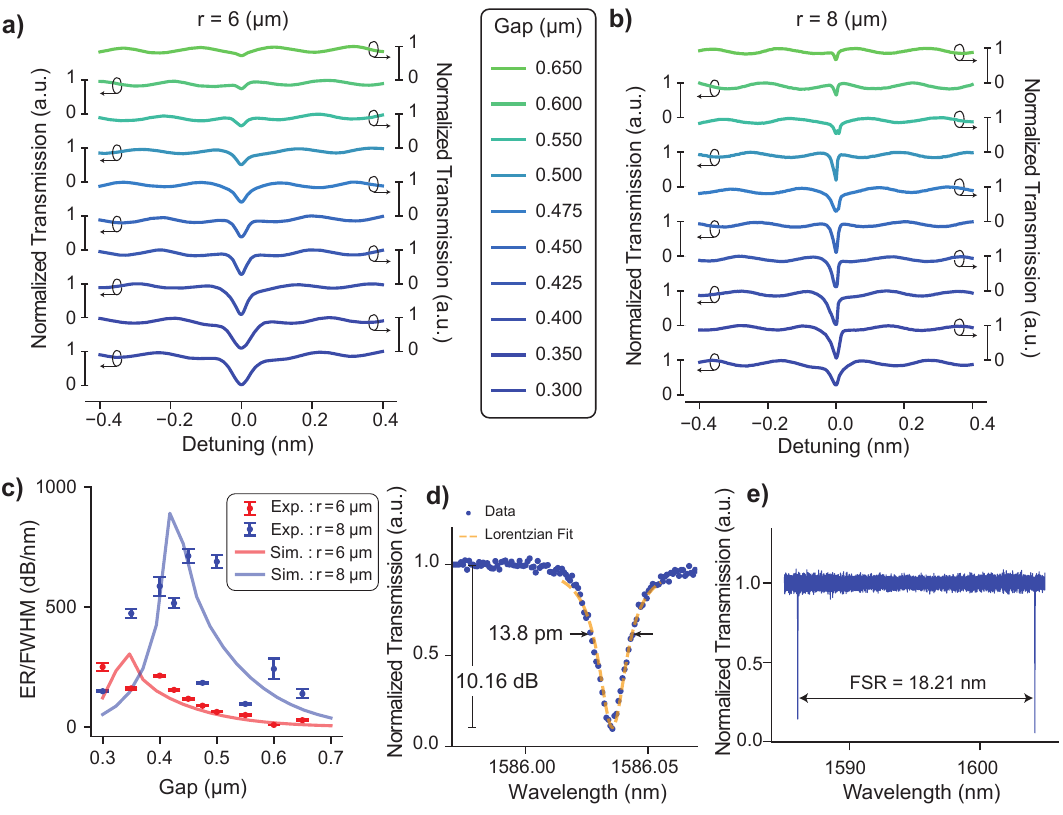}
\caption{Optical characterization of amorphous SiC single ring resonators. The through-port transmission spectra of single-ring resonators for different coupling gaps with a radius of \textbf{a)} \qty{6}{\micro\metre} and \textbf{b)} \qty{8}{\micro\metre}, respectively. \textbf{(a)} and \textbf{(b)} share the same legend. \textbf{c)} Comparison of the optical performance metric of the ring resonators with different coupling gaps and radii, together with the simulation results of the same geometries of the ring resonators. The ring with each radius' highest ER/FWHM ratio is selected for further ultrasound sensitivity analysis. \textbf{d)} Transmission spectrum of the PDMS-coated a-SiC micro ring resonator with $r =$ \qty{8}{\micro\metre}, $w_{a-SiC} =$  \qty{0.8}{\micro\metre},  and $gap =$ \qty{0.45}{\micro\metre}. FWHM and ER measurements are annotated on the graph, and the values are extracted from the Lorentzian fit to the data.  \textbf{e)} FSR measurement of the single ring resonator in \textbf{(d)} from a wide range of wavelength sweep. The data is normalized to the background oscillations in the sweep measurement. The sweep range includes two consecutive resonance notches of the same resonator.}
\label{figSingleRings}
\end{figure}

At the optical resonance wavelength, $\lambda_r$, of a ring resonator, its optical transmission $T$ shows a sharp resonance notch. To reach high sensitivity levels, it is desirable to maximize the slope $|\frac{\mathrm{d}T}{\mathrm{d}\lambda}|$ at the flank of this notch. This optimization can be achieved for circular ring resonators by varying the coupling gap between the ring resonator and the straight waveguide. In  Fig. \ref{figSingleRings}a and Fig. \ref{figSingleRings}b, the transmission response of ring resonators with different coupling gaps are plotted for ring radii of \qty{6}{\micro\metre}  and \qty{8}{\micro\metre}, respectively. The slope of the ring resonance flank determines the optical sensitivity in ultrasound sensing. Sensors with small full-width-half-maximum (FWHM) and high extinction ratio (ER, see Eq. (\ref{eq:ER})) provide a higher slope on the flank of the ring transmission response. We plot each resonator's ratio ER/FWHM in Fig. \ref{figSingleRings}c, as obtained from Lorentzian fitting, and for each ring radius select the coupling gap that results in the highest ER/FWHM for ultrasound characterization experiments. The error bars in Fig. \ref{figSingleRings}c quantify the fitting errors up to two standard deviations. 

\begin{figure}[!htb]
\centering
\includegraphics[width=1\textwidth]{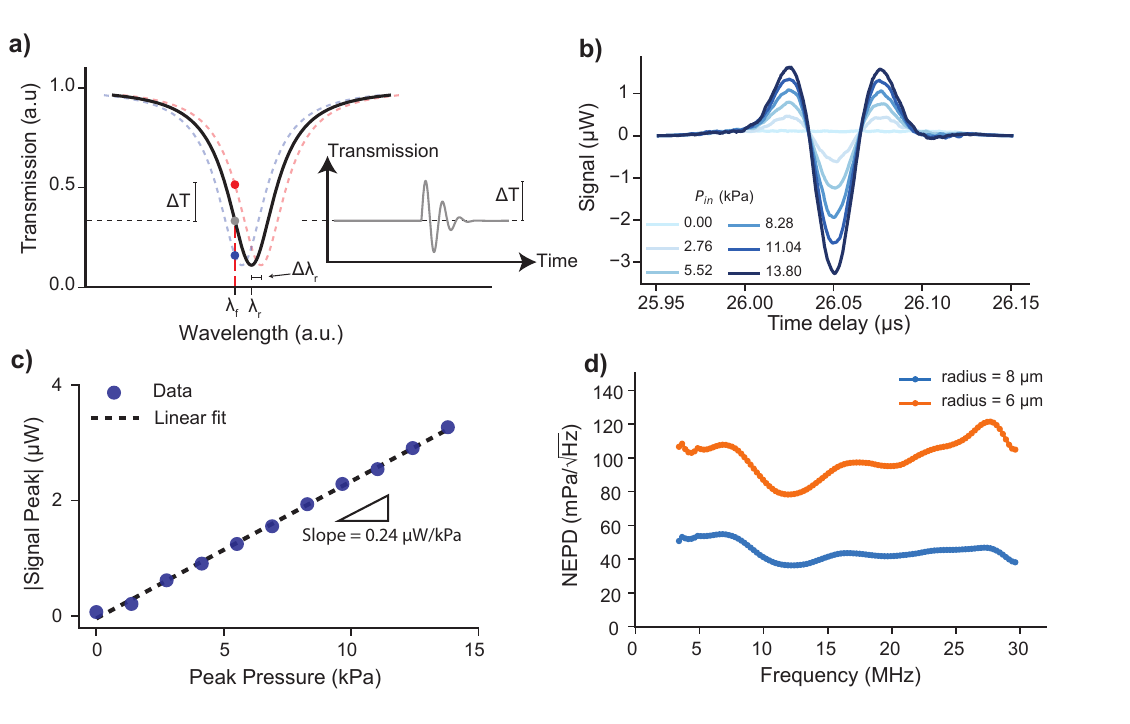}
\caption{Ultrasound sensitivity and NEPD characterization of amorphous SiC single ring resonators. \textbf{a)} Schematic of the read-out method by transmission output intensity monitoring when the laser wavelength is set to  $\lambda_f$ on the flank of the resonance. The resonance wavelength without applied ultrasound pressure and wavelength shift for an applied pressure is indicated as $\lambda_r$ and $\Delta \lambda_r$, respectively. For a certain ultrasound pulse, as indicated in the inset, $\Delta T$ indicates the maximum change in transmission of the ring resonance as a result of a $\Delta\lambda_r$ shift. The inset schematically demonstrates the recorded output signal in time. \textbf{b)} Recorded signals of the selected resonator under ultrasound pulses with varying pressures when the laser wavelength is set to the flank wavelength of the resonator. The signals are averaged 100 times and cut from a longer trace using a Tukey window. \textbf{c)} Plots the calibrated peak pressures against the absolute maximum of the measured signals, representing the linearity of the sensor in the range of applied pressures. The slope of the linear fit to the measured data is annotated in the plot area. \textbf{d)} NEPD of the sensor over the characterization bandwidth. The characterization bandwidth is limited by the bandwidth of the signal generated in the setup and received by the hydrophone. The lower frequency range limit is found by excluding the part of the full calibration spectrum where the signal amplitude reading from the hydrophone drops by 10 dB compared to its maximum. The high frequency range limit is set to the maximum calibration frequency of the hydrophone used in calibration measurements (see SI S4).}
\label{figSingleRingsUS}
\end{figure}

Figure \ref{figSingleRings}c shows that for rings with a radius of \qty{6}{\micro\metre}, the ratio ER/FWHM is in general lower than that of the rings with \qty{8}{\micro\metre} radius and that the optimal coupling gap occurs at smaller gap distances. We verify this observation with finite difference time domain simulations in Fig. \ref{figSingleRings}c. To obtain the shown agreement between simulation results and experiments, the ring resonators' one round-trip amplitude losses are set to \qty{3}{\decibel\per\cm} and \qty{9}{\decibel\per\cm} for \qty{8}{\micro\metre} and \qty{6}{\micro\metre} ring radii, respectively. In Fig. \ref{figSingleRings}d, the transmission spectrum measurement of the ring resonator with the highest ER/FWHM in our study is shown in detail. The resonator has a FWHM of \qty{13.8}{\pico\metre} corresponding to a Q factor of above $1.1\times10^5 $ and an ER of \qty{10.16}{\decibel}.

A high finesse in parallelized ring resonator arrays for ultrasound detection is important because it is directly related to the number of rings that can be simultaneously coupled and individually read out via a single bus waveguide. Figure \ref{figSingleRings}e provides a measurement of the free spectral range (FSR) of the ring with a radius of \qty{8}{\micro\metre}. The distance between the two consecutive resonance modes in our sweep range is measured as \qty{18.21}{\nano\metre}. According to this, the finesse of the resonator is calculated to be around 1300. 

The schematic in Fig. \ref{figSingleRingsUS}a demonstrates the measurement principle of ultrasound sensing with optical microring resonators. The ring resonator has a resonance notch at the specific wavelength, $\lambda_r$. The impinging ultrasound signal changes the resonance wavelength of the microring resonator by $\Delta\lambda_r$. By interrogating the transmission of the ring resonator at a flank wavelength, $\lambda_f$, the ultrasound pressure-induced wavelength shifts can be converted to time-variations in optical transmission as is indicated by $\Delta T $ in  Fig. \ref{figSingleRingsUS}a. The inset plots the transmission change over time due to the ultrasound signal received. 

We select the ring resonators with the highest figure of merit (Fig. \ref{figSingleRings}c) and characterize them under different ultrasound peak pressures. For these characterization experiments, the ultrasound transducer is excited with a Gaussian pulse at a central frequency of \qty{19.6}{\mega\hertz}. The recorded signals as a response to this pulse from \qty{8}{\micro\metre} radius ring for different calibrated peak pressures are plotted in Fig. \ref{figSingleRingsUS}b. The resonator's peak response increases as the pulse's peak pressure increases. Figure \ref{figSingleRingsUS}c plots the absolute maximum value of the measured signal peaks given in Fig. \ref{figSingleRingsUS}b, demonstrating the linearity of the sensor's acoustic response at this pressure interval. The slope of the curve is found to be \qty{0.24}{\micro\watt\per\kilo\pascal}. The pressure calibration procedure for the responsivity measurements is detailed in the Supplementary Information (SI) Section 4. The spectral responsivity , which is on average \qty{9.7}{\milli\V\per\kilo\pascal} (see Fig. S5c), is determined from the time domain signals by applying the Fourier transform. For this, we use the response of the sensor and a calibrated hydrophone's response to the same ultrasound signal with peak input pressure of \qty{13.8}{\kilo\pascal}. This spectral response is further used to evaluate the sensor's NEPD (see SI Section 5).

To determine the NEPD of the selected sensors, as shown in Fig. \ref{figSingleRingsUS}d, we divide the noise amplitude spectral density of the resonator for each frequency by the spectral responsivity of the resonator. The noise amplitude spectral density is obtained from a time trace recorded without any form of averaging and without applying ultrasound pressure at the same flank wavelength and laser power as the measurements in Fig. \ref{figSingleRingsUS}b (see also \cite{westerveld_sensitive_2021} for the detailed methodology).

\begin{figure}[!htb]
\centering
\includegraphics[width=1\textwidth]{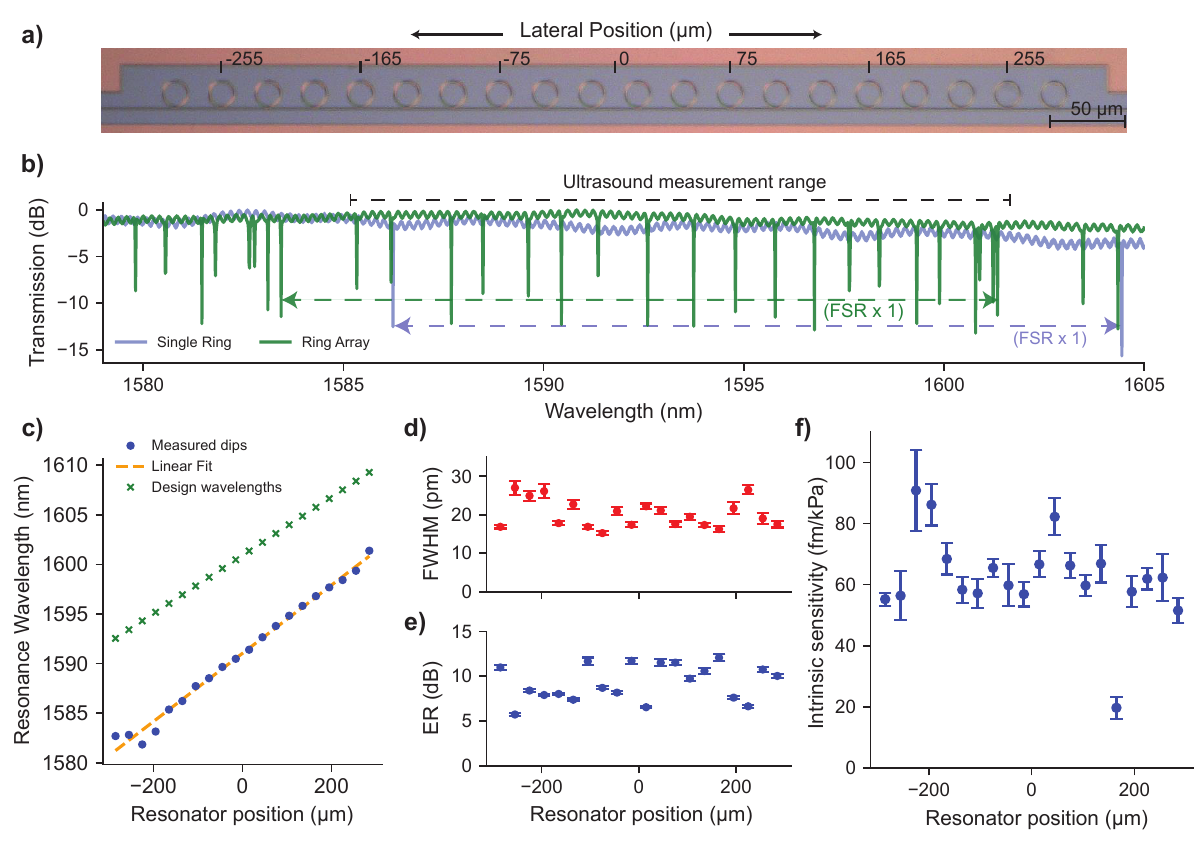}
\caption{Intrinsic sensitivity and optical characterization of an array of ring resonators coupled to single bus waveguide. \textbf{a)} Microscope image of the twenty rings used in ultrasound characterization experiments. The radius of each ring is slightly increasing from left to right of the image. The pitch of the array is \qty{30}{\micro\metre}. The lateral axis is annotated, and a ruler is provided along the lateral axis over the array.  \textbf{b)} Measured spectra of the single ring resonator and the linear array of 20 a-SiC ring resonators. FSR and the range of ultrasound measurement notches are indicated with annotations. \textbf{c)} The resonance wavelengths of the ring resonators are plotted against their lateral position. Resonator positions are determined using the ultrasound sensing and imaging experiments. A linear fit to measured resonance wavelengths is also plotted. Resonators follow a linear trend except for the first four resonators of the array. Intended design wavelengths of the resonators are also provided in the same plot.  \textbf{d)} and \textbf{e)} Full-width half maximum (FWHM) and extinction ratio (ER) of each resonator in the array. Error bars in each sub-figure represent 2-sigma variation over the 20 different measurements of the same array with 20 resonators. \textbf{f)}  Intrinsic ultrasound sensitivity of each resonator in the array. This sensitivity is a measure independent of the resonator's linewidth and coupling effects. }\label{figResonatorArrays}
\end{figure}

In Fig. \ref{figResonatorArrays}a, a microscope image of an array of 20-ring resonators, coupled to a single straight bus waveguide, is shown. Each ring has a radius of approximately \qty{8}{\micro\metre} and gap distance of \qty{450}{\nano\metre} since these parameters gave the best NEPD performance in the single ring optical characterization experiments. The pitch (center to center distance) between the resonators is \qty{30}{\micro\metre}, and each ring has a slightly different designed radius starting from \qty{8.000}{\micro\metre} up to \qty{8.116}{\micro\metre}, gradually increasing from the left of the image to the right with a step size of \qty{5.8}{\nano\metre}. The waveguide width of the bus and ring waveguide is \qty{800}{\nano\metre}. 

\begin{figure}[]
\centering
\includegraphics[width=0.92\textwidth]{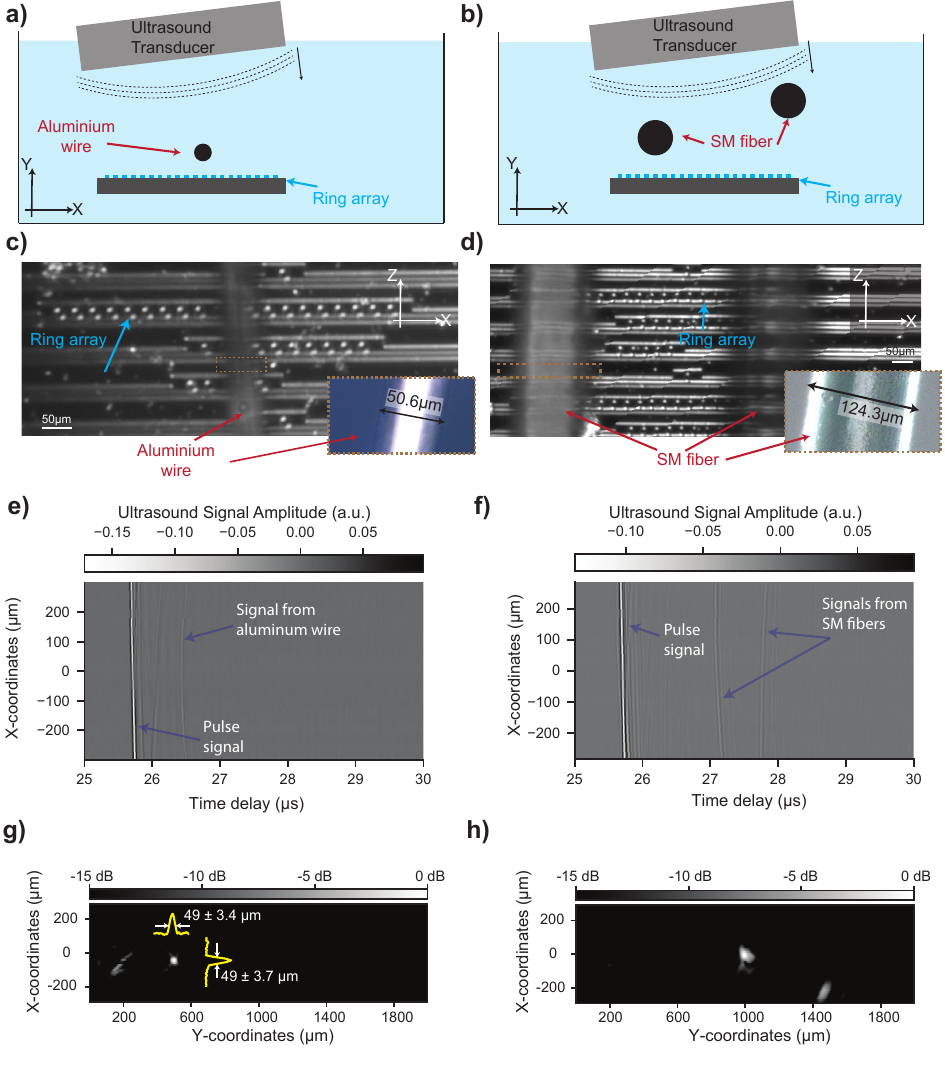}
\caption{ Ultrasound tomography imaging demonstration with the array of a-SiC ring resonators. \textbf{a)} and \textbf{b)} Schematic views of the ultrasound tomography experiments for \qty{50}{\micro\metre} diameter aluminum wire and two of \qty{125}{\micro\metre} diameter single mode fibers, respectively. The objects are in the water tank between the tilted ultrasound transducer and the array of rings at different heights. They are shown as a black circle, which corresponds to their cross-section. Inset plots the excitation signal of the ultrasound transducer to generate the ultrasound pressure wave.\textbf{c)}  Microscopy image of the ring array and the aluminum wire. The optical focus is adjusted to the chip plane. The inset shows the microscopy image of only the aluminum wire with higher magnification.  \textbf{d)} Microscopy image of the ring array and the two single-mode (SM) optical fibers. The optical focus is adjusted to the chip plane. The inset shows the microscopy image of one of the optical fibers with higher magnification. \textbf{e)} and \textbf{f)} Time traces of the signals in the a-SiC ring resonator array registered during the tomography experiments. The initial pulse arrives at different times at each sensor location and is correlated with the tilt angle. The signals from the aluminum wire and the SM fibers are indicated with arrows and text on the time traces.
\textbf{g)} Reconstructed image using frequency domain image reconstruction algorithm by using the time traces shown in \textbf{(e)}. \textbf{h)} Reconstructed image using frequency domain image reconstruction algorithm using the time traces shown in \textbf{(f)}. The FWHM measurements of the wire in \textbf{(g)} and fibers in \textbf{(h)} are extracted from the yellow curves that are overlayed on the image by fitting a Gaussian function.}\label{figImage}
\end{figure}

Figure \ref{figResonatorArrays}b plots the laser wavelength sweep measurement for the array of 20 a-SiC ring resonators and for a single ring resonator. By design, the individual responses of all 20 rings are distributed within a single free spectral range (FSR). Each resonance notch within the FSR corresponds to a different ring resonator in the array, allowing us to read-out each ring passively—without requiring resonance wavelength tuning mechanisms such as heaters. Ultrasound measurements are performed using the 20 consecutive resonances as indicated in Fig. \ref{figResonatorArrays}b. Due to random variations in fabrication, a sorting methodology is applied after the ultrasound measurements to match the positions of the rings in the chip with their resonance wavelengths in the spectrum sweep. The details of this sorting are provided in SI S6.

Extracted resonance wavelengths of the ring resonators are plotted in Fig. \ref{figResonatorArrays}c from the sweep measurement of the ring array against their determined positions from ultrasound measurements. The standard deviation of the difference between measured resonance wavelengths from the linear fit is calculated to be \qty{0.584}{\nano\metre} with a maximum difference of \qty{1.45}{\nano\metre}.  Most of the resonators align with a linearly increasing trend except the first four resonators, which may result from a local fabrication variability. Moreover, we observe significant non-uniformity of the ring notch responses in the linear array in terms of the FWHM and the ER of each ring resonator. This suggests that there is a significant variation of the coupling between the bus and the ring waveguide. Figure \ref{figResonatorArrays}d and Fig. \ref{figResonatorArrays}e plot the results of FWHM and ER of each ring response in the array as a result of a Lorentzian fitting to their notch responses. We also determine the intrinsic sensitivity ($S_{int}=\mathrm{d} \lambda_r/\mathrm{d}  P$) of the resonators (see Fig. \ref{figResonatorArrays}f) using the methodology described in \cite{erdogan_PDMS,zarkos_fully_2023}. We analyze the factors that determine the intrinsic sensitivity in detail in SI S2. We observe that the intrinsic sensitivity distribution over the array has a mean of \qty{62}{\femto\metre\per\kilo\pascal} and is between \qty{19} and \qty{95}{\femto\metre\per\kilo\pascal}. A theoretical calculation is provided in the SI S2 and SI Table 1, resulting in $S_{int}=$ \qty{78}{\femto\metre\per\kilo\pascal}, in good agreement with the measured values. The intrinsic sensitivity is governed by the pressure dependence of the resonance wavelength shift and not by the shape of the resonance notch. This difference in governing physics can explain the significant variations in values for FHWM, ER, and $S_{int}$ in Figs. \ref{figResonatorArrays}d, \ref{figResonatorArrays}e and \ref{figResonatorArrays}f, while the variations in $S_{int}$ seem to be relatively small.

To test the suitability of the array of a-SiC ring resonators for ultrasound tomography, we performed imaging experiments for two scenarios. Schematic cross-sectional views of the two experiments are given in Fig. \ref{figImage}a and Fig. \ref{figImage}b. For the first experiment, an aluminum wire with a measured diameter of \qty{50.6}{\micro\metre} is placed in the water tank between the ring array and the ultrasound transducer, as shown in Fig. \ref{figImage}c. Two single-mode fibers with a diameter of \qty{124.3}{\micro\metre} are placed at different heights for the second experiment, as shown on the micrograph in Fig. \ref{figImage}d. We use custom 3D-printed holders for these imaging experiments (see SI Fig. S1). The ultrasound pulse generated by the ultrasound transducer at $t=0$ travels in the negative Y direction, is reflected from the chip, then travels in the positive Y direction, reflects from the sample (aluminium wire or SM fiber), and then reaches the a-SiC array on the silicon chip for the second time. By analyzing both the first and second ultrasound pulses with the a-SiC array of microring resonators, we create an ultrasound image of the sample. We use this pulse-echo method to demonstrate the imaging performance of the sensor under a representative imaging scenario for the targeted ultrasound imaging applications.

The recorded time traces for both imaging scenarios (single aluminum wire or two SM fibers) are plotted in Fig. \ref{figImage}e and Fig. \ref{figImage}f. The time response of each of the 20 rings is included at the corresponding X-coordinate of the ring. We observe the initial arrival of the pulse to the ring array and the reflection signals from the aluminum wire and the SM fibers following the pulse signal after a time delay. The time-of-flight measurements agree well with k-wave simulations presented in Fig. SI S2c and S3c.

The resulting tomography images captured by a-SiC array are provided in Fig. \ref{figImage}g and Fig. \ref{figImage}h for the aluminum wire and the two SM fibers, respectively. The details about the image reconstruction algorithm are explained in the Methods section. The yellow curves in Fig. \ref{figImage}g. are extracted from the image data at the location of the samples along the lateral and axial directions. The FWHMs of the single wire with \qty{50.6}{\micro\metre} diameter in Fig. \ref{figImage}g along the X and Y directions are measured by using a Gaussian fit to the image signal and are found to be \qty{49}{} $\pm$ \qty{3.7} {\micro\metre} and \qty{49}{} $\pm$ \qty{3.4}{\micro\metre}, respectively. Uncertainty in the measured values is determined from the SNR of the fitted image data itself. The same procedure is also applied to the image of the SM fibers, and the FWHM of the image signals are \qty{96}{} $\pm$ \qty{7.6}{\micro\metre} and \qty{161}{} $\pm$ \qty{35.2}{\micro\metre} in Y direction;  and \qty{89}{} $\pm$ \qty{11.5}{\micro\metre} and \qty{139}{} $\pm$ \qty{19}{\micro\metre} in X direction. The measurements agree with the sample diameter for the single aluminum wire. Although the results for SM fibers are found to be close to the actual diameter of the fibers, we attributed the difference to the imaging distance, hence the numerical aperture of the array at this depth of imaging. The numerical aperture of the array in Fig.\ref{figImage}g. is 0.6 at the depth of the aluminum wire. On the other hand, in Fig.\ref{figImage}h., the numerical aperture becomes 0.28 and 0.2 for the SM fibers at depths of \qty{0.95}{\milli\metre} and \qty{1.3}{\milli\metre}, respectively. Thanks to the high finesse of a-SiC ring resonator platform, the accuracy of the measurement of the diameter and the imaging quality can further be improved by increasing the number of rings in the array, so that the numerical aperture will be higher for the same depth values.

\section*{Discussion}\label{sec3}

\subsection*{Performance of the array of integrated optical ultrasound sensors for imaging}\label{subseccomparison}

We can theoretically compare the NEPD performance of the a-SiC ultrasound sensors to the widely used piezoelectric elements for ultrasound detection in photoacoustic imaging using the theoretical NEPD estimation of piezo elements in photoacoustic imaging \cite{piezo_NEPD_theory}. The NEPD of piezoelectric elements can be estimated using the equation: 

\begin{equation}
    \text{NEPD}(f)_{piezo} = \sqrt{\frac{F_n k T Z_a}{A \eta(f)}}
    \label{nepdpiezo}
\end{equation}

where $F_n$ is the noise factor of the preamplifier, $k$ is the Boltzmann constant, $T$ is the temperature, $Z_a$ is the acoustic impedance of the medium, $A$ is the area of the piezoelectric element, and $\eta(f)$ is the detector efficiency. Considering the conservative calibration range in our experiments, the high cut-off frequency is \qty{30}{\mega\hertz}. For an ultrasound imaging array of transducers designed for this high-cut-off frequency, a half-wavelength matched piezoelectric transducer element should be of a size of \qty{25}{\micro\metre} to avoid spatial aliasing. According to Eq. (\ref{nepdpiezo}) and taking typical efficiency values for broadband transducers, $\eta{(f)}$, in the range between $0.01$ and $0.001$, the piezoelectric elements of this size theoretically have a NEPD between \qty{45}{\milipascal\per\sqrt{Hz}} and \qty{131}{\milipascal\per\sqrt{Hz}}, respectively. On the other hand, a-SiC ultrasound sensor with a diameter of \qty{16}{\micro\metre} performed with an average NEPD of \qty{45}{\milipascal\per\sqrt{Hz}} on average in the range of our calibration (see SI Section NEPD). Since we consider a theoretical piezoelectric element size matching the half wavelength of the high-cut-off frequency of the ultrasound signal, there is no other metric to be introduced for this theoretical calculation. This is because there is no advantage for a piezoelectric transducer in being smaller than half of the smallest acoustic wavelength in the generated imaging signal \cite{garrett_acoustic_2021}. 

The measured NEPD of our a-SiC platform is over \qty{20} times better than that of the smallest hydrophones (\qty{1000}{\milipascal\per\sqrt{Hz}} at \qty{40}{\micro\metre} diameter \cite{pa40um}), while occupying only one-fourth of their chip area. Furthermore, to our knowledge, the smallest demonstrated piezoelectric element size in a piezoelectric array of detectors is \qty{100}{\micro\metre}, which makes them unsuitable for imaging at these higher frequencies or in near-field imaging scenarios without being affected by Nyquist sampling errors or compromising their sensitivity. Thus, a-SiC integrated photonic ultrasound sensors offer better sensitivity than piezoelectric elements without compromising the miniaturization in ultrasound sensing for high-frequency ultrasound signals.

To evaluate the array of a-SiC photoacoustic tomography sensors in comparison to the state-of-the-art of integrated photonic ultrasound sensor arrays, Table \ref{tab:comparison} lists the performance parameters of only the array of optical sensors designated for PAI applications.
 
Table \ref{tab:comparison} indicates that the demonstrated a-SiC resonator array showcases metrics that are competitive in all metrics while showing best-in-class performance in terms of finesse, pitch, and parallel detector count. It is of interest to assess to what extent this performance is due to the material properties of the a-SiC core material. Therefore, we compare the theoretical performance of the a-SiC waveguides from this work to that of Si, SiN, and ChG waveguides used in ultrasound sensing from the literature in SI Table 1. According to the SI Table 1, the intrinsic sensitivity performance of our a-SiC platform is theoretically in between Si and SiN waveguides, thus providing a good balance between miniaturization and high sensitivity with $S_{int}$ twice higher than that of Si waveguides. 

Besides refractive index differences, another important aspect is the sensor's propagation and coupling losses, which determine its FWHM and finesse. From the low FWHM and high finesse in Table \ref{tab:comparison}, the performance of the a-SiC waveguide platform boasts a very good loss performance compared to silicon and polystyrene devices. Several factors can contribute to this, which include high fabrication quality (layer roughness and waveguide uniformity), sufficiently large ring radius, optimized coupling gap (Fig. \ref{figSingleRings}), and low optical absorption of the a-SiC material and cladding layers. The large free spectral range and low propagation loss of the a-SiC devices enable including many ring resonators on the same bus waveguide that can be read out individually or simultaneously \cite{pan_parallel_2023}. Finally, the low growth temperature of the a-SiC of 150$^\circ$C is a clear advantage, especially when integrating them on platforms that cannot withstand high temperatures, like flexible substrates or the backend of complementary metal-oxide-semiconductor (CMOS) wafers.

\begin{table}[!htb]
\addtolength{\tabcolsep}{0em}
\begin{tabularx}{460pt}{@{}lcccccccccc@{}}
\toprule
Resonator  &
\begin{tabular}[c]{@{}c@{}}Platform\end{tabular} &
  \begin{tabular}[c]{@{}c@{}}FWHM\\ (\unit{\pico\metre})\end{tabular} &
  \begin{tabular}[c]{@{}c@{}}Finesse\\ \end{tabular} &
  \begin{tabular}[c]{@{}c@{}}$S_{int}$\\ (\unit{\femto\metre\per\kilo\pascal})\end{tabular} &
  \begin{tabular}[c]{@{}c@{}}Aperture \end{tabular} &
  \begin{tabular}[c]{@{}c@{}} Pitch\\ ($\mu m$)\end{tabular} &
  \begin{tabular}[c]{@{}c@{}}Parallel\\detector\\count\end{tabular}&
  \begin{tabular}[c]{@{}c@{}}Maximum\\film growth\\temperature \\(\unit{\degreeCelsius}) \end{tabular}&
  \begin{tabular}[c]{@{}c@{}}$\frac{S_{int}}{FWHM}$ \\ (1/MPa) \end{tabular}\\ \midrule
  $\pi$-BG \cite{hazan_miniaturized_2022} &SOI & 41.1 & N/A            & 188   & $265 \times 0.5$ \unit{\micro\metre\squared} & 130\** & $5$ & 1100 & 4.58    \\ 
\midrule 
 Ring \cite{westerveld_sensitive_2021}      &  SOI-MEMS  & 83  &  \textbf{205}   &  \textbf{40000}      &20 \unit{\micro\metre}   & \textbf{30}  & $10$ & 1100 & \textbf{481} \\ 
 \midrule
Ring \cite{maxwell_polymer_2008}         &  PS & 305 & -   &      200     & 100 \unit{\micro\metre}  & 200& 4 & \textbf{180} & 0.65 \\ 
Ring \cite{zarkos_fully_2023}           & SOI & 130 & 147$^{\dagger\dagger}$      &  40     &\textbf{10} \unit{\micro\metre} & 250$^{**}$ & 8 & 1100 & 0.307   \\ 
Ring  \cite{pan_parallel_2023}        & ChG & \textbf{2.6}$^{*}$ & - &\textbf{410}$^\dagger$   &40 \unit{\micro\metre}   & 400 & \textbf{15} & 350 \cite{Xia_anneal_ChG} & \textbf{157}  \\
Ring& a-SiC & \textbf{13.8} & \textbf{1320}  &  78     &\textbf{16} \unit{\micro\metre}  & \textbf{30}   & \textbf{20} & \textbf{150} & 5.65  \\ \bottomrule
\end{tabularx}
\caption{Properties of the reported array of integrated photonic ultrasound sensors. Best 2 performance metrics in boldface. Parameters that couldn't be found or calculated are left blank. SmartCut processing technology is assumed for SOI platforms for process temperature column \cite{Bruel_smartcut_1997}. PS: polystyrene, SOI: silicon on insulator, N/A: Not applicable.\\ \** Extracted from the published figure of the reference article.
\\ \*** Includes integrated debugging and tuning interface blocks on the chip.
\\ \* $\dagger$ Calculated from parameters in \cite{BaoReview} for a similar material.
\\ \* $\dagger\dagger$ Calculated using reported parameters and the equation $Finesse=\frac{FSR}{FWHM}$ and $FSR=\lambda^2/n_gL$ where lambda is taken as $1550$ nm, $n_g$ is the calculated group index of the TE0 mode for an SOI strip waveguide with FEM, and $L$ is the circumference length of the ring resonator.}
\label{tab:comparison}
\end{table}

In conclusion, we have presented a linear array of twenty a-SiC ring resonators utilizing wavelength division multiplexing, demonstrating its suitability for high-sensitivity ultrasound sensing and acoustic tomography applications. Our results highlight the advantages of a-SiC as a waveguiding material over traditional platforms, featuring low FWHM and high finesse, while also demonstrating its miniaturization potential via its small aperture and pitch. Importantly, we discuss the potential combination of the a-SiC waveguides with CMOS electronics, thanks to the film growth temperatures as low as \qty{150}{\degreeCelsius}. These features of a-SiC enable the development of sensitive sensor arrays without compromising resonator size, leading to improved spatial resolution and contrast for the next generation of advanced PAT applications that require miniaturized ultrasound sensors, such as living mouse brain imaging or intra-vascular imaging with catheters.
\newpage
\section*{Methods}\label{sec4}
\subsection*{Fabrication of Amorphous Silicon Carbide Ring Resonators}\label{subsec1}
Fabrication of the ultrasound sensors follows the same fabrication steps published earlier \cite{bruno_acs} at a low deposition temperature of 150$^\circ$C. For our application, the final silicon dioxide top cladding is substituted by an elasto-optic polymer, PDMS, with a measured thickness of around \qty{5}{\micro\metre}. This is achieved by spin coating a PDMS mixing ratio of 1:30, and the detailed steps of the coating process are published earlier in \cite{erdogan_PDMS}.

\subsection*{Ultrasound Characterization Setup}\label{subsec2}

We used the same setup published earlier \cite{erdogan_PDMS} to perform the ultrasound characterization of the rings. In summary, we couple the laser light using two grating couplers; we submerge everything underwater as depicted in Fig. \ref{figConcept}a. We align the ultrasound source and launch Gaussian pulses directed to the sensor. By varying the peak-to-peak amplitude, we can characterize the sensitivity of the individual sensors and record the ultrasound wavefield by tuning the laser to the flank wavelength of each resonator in the array. The same setup is used for imaging experiments with an additional sample holder attached to the chip holder; sample holders for imaging are shown in SI S1. 

\subsection*{Ultrasound image reconstruction}\label{subsec3}

We use the sequentially recorded ultrasound data from each resonator in the linear array and the frequency domain reconstruction method to perform image reconstruction. In general, beamforming is performed by applying a 2D fast Fourier transform (FFT) to the received echoes, interpolating the data onto an acoustic dispersion grid, and then applying a 2D inverse FFT (IFFT) to reconstruct the signals. This method, described in \cite{Jian_beamform}, enables Fourier-domain beamforming when using a plane wave for insonification.
In more details, since the plane wave transmitted is angled, received data is first shifted in the frequency domain by multiplying by the propagator $e^{-i\omega\tau_{ch}}$ where $\tau_{ch}$ represents the individual channel delay steering the beam at the given angle $\Theta$, $\omega$ is the angular frequency, $\omega=2\pi f$,  here $f$ is taken as the center frequency of the transducer. Data is then interpolated in time to increase the sampling frequency and apodized by a 2D Tukey window as well as zero-padded (axially and laterally) to ensure smooth and sufficient frequency bins for further interpolation in the Fourier domain when applying a 2D FFT. Data is then remapped in the frequency domain into the object grid through complex shifting $ k = \frac{k_y^2 + k_x^2}{2k_y \cos(\Theta) + 2k_x \sin(\Theta)} $, where $ k_x = \frac{2\pi}{\text{pitch}}$ (with "pitch" being the element spacing), and  $k_y = \frac{2\pi}{F_{\text{sampling}}}$. The beamformed image is then obtained after applying an IFFT.

\subsection*{Optical characterization of ring resonators}\label{subsec4}
To characterize the spectral response of a-SiC microring resonators, FWHM and ER of the single-ring resonators were extracted by fitting a Lorentzian function to the measured transmission spectrum. The measured transmission spectrum near the resonance was modeled and fitted using the Lorentzian function:
\begin{equation}
    T(\lambda) = \frac{a_1}{(\lambda - \lambda_0)^2 + \left(\frac{\gamma}{2}\right)^2} + c
    \label{transmission}
\end{equation}
where $\lambda$ is the wavelength, $a_1$ is the amplitude scaling parameter (it has a negative value), $\lambda_0$ is the resonance wavelength, $\gamma$ is the FWHM of the resonance dip, and $c$ is the baseline offset. Using the fit result, the ER was calculated as: 
\begin{equation}
    \text{ER}_{\mathrm{dB}} = 10 \cdot \log_{10} \left(  \frac{c}{\min(T)}  \right)
    \label{eq:ER}
\end{equation}

The loaded quality factor of the resonance was computed as: $Q = \frac{\lambda_0}{\text{FWHM}}$. This analysis was applied to all transmission spectra after low-pass filtering and normalization.
\backmatter

\bmhead{Supplementary information}

This article has an accompanying Supplementary Information section.

\bmhead{Acknowledgments}
I.E.Z. acknowledges funding from the European Union’s Horizon Europe research and innovation program under grant agreement No. 101098717 (RESPITE project) and Dutch Research Council  (NWO) OTP COMB-O project (18757). R.T.E. and P.G.S. acknowledge support from the PhoQuS-T project. The project (23FUN01 PhoQuS-T) has received funding from the European Partnership on Metrology (Funder ID: 10.13039/100019599), co-financed from the European Union’s Horizon Europe Research and Innovation Programme and by the Participating States.

\section*{Declarations}

\begin{itemize}
\item Funding: Funding declarations are provided in acknowledgment section. 
\item Conflict of interest/Competing interests: The authors declare no conflict of interest.
\item Ethics approval and consent to participate: Not applicable
\item Consent for publication: All authors read and agreed to the final version of the manuscript to be submitted.
\item Data, Materials, Code availability: Data, Materials, and Code are available from the corresponding author on reasonable request.
\item Author contributions:
R.T.E., G.J.V. and P.G.S. conceptualized the study and designed the experiments. B.L.R. and R.T.E. fabricated the devices. R.T.E. conducted the simulations, experiments, and data analysis/visualization. R.T.E. and S.I.R. implemented the ultrasound image reconstruction algorithm and visualization. 
R.T.E. and P.G.S. wrote the original draft. All authors contributed to the manuscript review and editing. W.J.W, S.I.R, I.E.Z, and P.G.S provided the resources and funding. G.J.V and P.G.S. supervised the study.

\end{itemize}

\bibliography{sn-bibliography.bib}

\clearpage


\section*{Supplementary material}

# Supplementary Information: Amorphous Silicon-Carbide Photonics for Ultrasound Imaging

R. Tufan Erdogan[1*], Bruno Lopez-Rodriguez[2], Wouter J. Westerveld[1], Sophinese Iskander-Rizk[1], Gerard J. Verbiest[1], Iman Esmaeil Zadeh[2], Peter G. Steeneken[1*]

[1]Department of Precision and Microsystems Engineering, TU Delft, The Netherlands.
[2]Department of Imaging Physics (ImPhys), TU Delft, The Netherlands.

*Corresponding author(s). E-mail(s): r.t.erdogan@tudelft.nl; p.g.steeneken@tudelft.nl;
Contributing authors: b.lopezrodriguez@tudelft.nl; w.j.westerveld@tudelft.nl; s.iskander-rizk@tudelft.nl; g.j.verbiest@tudelft.nl; i.esmaeilzadeh@tudelft.nl;

## Contents

# 1 Sample holders for ultrasound imaging experiments

For imaging experiments we design and 3D print two type of holders. In Fig. S1a and Fig. S1b, photographs of printed parts are provided. The parts are designed such that the wires can be attached to the side of the holder and the holder can easily be attached to the aluminum chip holder as shown in Fig. S1c and Fig. S1d. The height of the the aluminum wire was approximately 500 µm, and two fibers were approximately at the height of 1 mm and 2 mm from the surface of the chip. However, due to the attractive forces when submerged under water the fiber that is placed at 2 mm distance from the chip deformed and got closer to the lower one. This was also evident from the photographs taken after the experiments that is shown in Fig. S1d.

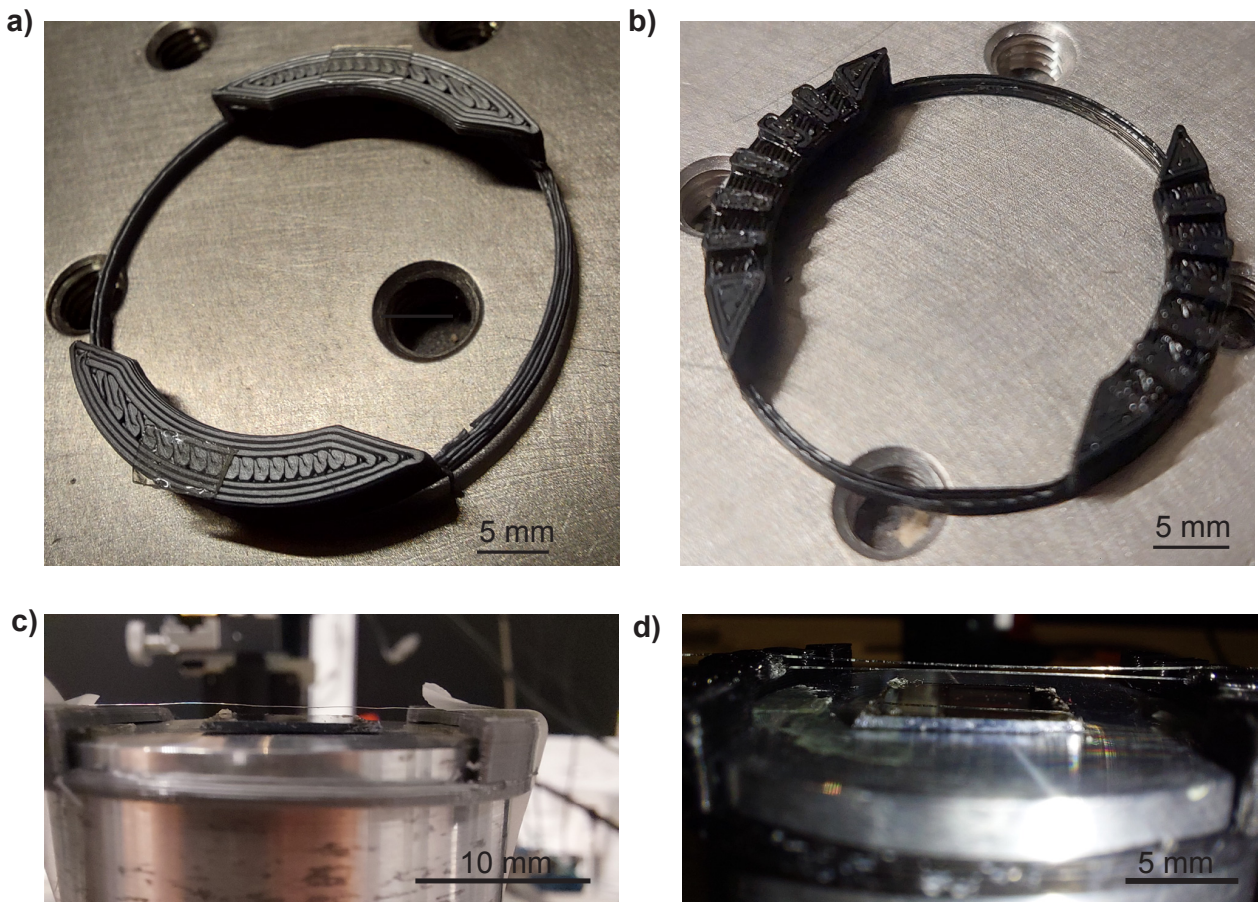

**Fig. S1** Sample holders and fixture for the imaging experiments. **a)**, and **b)** aluminum wire holder and SM fiber holder, respectively. **c)**, and **d)** Placement of the holders shown in **(a)**, and **(b)** to the aluminum chip stage after the aluminum wire and SM fibers are attached and used in experiments.

## 2 Ultrasound sensitivity of the ring resonators

The voltage at the output, $V_{out}$, of the photoreceiver connected to the output of the ring resonator through-port can be written as:

$$V_{out}(\lambda_f, \partial\sigma) = \psi P_{out} T(\lambda_f, \partial\sigma) = \psi P_{out} T(\lambda_f) + \left.\frac{\partial V_{out}}{\partial \sigma}\right|_{\lambda_f} \partial\sigma \tag{1}$$

where $\psi = R_{PD}G_{TIA}$ is a constant composed of the photodiode responsivity and the gain of the trans-impedance amplifier and $P_{out} = \eta_{gra}P_{laser}$ is the output power of the light in which $\eta_{gra}$ refers to total coupling efficiency of the two grating couplers. The output voltage depends on the output light power, $P_{out}$, and the transmission function of the ring resonator at the flank wavelength ($\lambda_f$), $T(\lambda_f, \partial\sigma)\big|_{\lambda_f}$. This transmission function can be expanded to include a perturbation term of pressure, $\partial\sigma$, that accounts for changes in the ultrasound pressure.

To quantify the total sensitivity of the sensor, $S_{tot}$, we can express it in terms of two distinct contributions relevant to the sensors' design and waveguide properties, $S_{ph}$ and $S_{int}$, and other constants regarding the experimental setup:

$$S_{tot} = \left.\frac{\partial V_{out}}{\partial \sigma}\right|_{\lambda_f} = P_{out}\psi S_{ph} S_{int} \tag{2}$$

where $P_{out}$ and $\psi$ are the experimental setup related constants, $S_{ph}$ is the sensitivity of photonic transduction of the transmission amplitude monitoring method, and $S_{int}$ is the intrinsic sensitivity of the resonator as a change in wavelength to the applied pressure. Accordingly, the total sensitivity can be rewritten in terms of the derivative of the transmission function at the flank wavelength and resonance wavelength shift due to pressure:

$$S_{tot} = P_{out}\psi \left.\frac{\partial T}{\partial \lambda_{res}}\right|_{\lambda_f} \frac{\partial \lambda_{res}}{\partial \sigma} \tag{3}$$

The third term on the right side of Eq. 3, represents the photonic contribution to sensitivity:

$$S_{ph} = \left.\frac{\partial T}{\partial \lambda_{res}}\right|_{\lambda_f} \sim \frac{1}{FWHM} \sim \frac{Q}{\lambda_{res}} \tag{4}$$

the reciprocal of full width at half maximum (FWHM) characterizes the slope of the resonance transmission spectrum for high-quality factor resonators operated at the flank wavelength, and $Q$ is the resonator's quality factor. The propagation losses in the ring resonator and the coupling between the bus waveguide and the ring waveguide determine the Q-factor of the resonator.

The last term on the right side of the Eq. 3, captures the intrinsic sensitivity of the ring waveguide, which quantifies the shift in resonance wavelength due to applied pressure. Since the core materials we consider here have high Young's modulus and the applied pressure amplitudes of photoacoustic signals are small, the

3

deformation of the waveguide core and change in the ring circumference can be ignored. In this case, only photoelastic changes in the refractive index of materials are considered:

$$S_{int} = \frac{\partial \lambda_{res}}{\partial \sigma} = \frac{\partial \lambda_{res}}{\partial n_{eff}} \frac{\partial n_{eff}}{\partial \sigma} \tag{5}$$

Expanding the dependence of the effective index on different material components of the waveguide:

$$S_{int} = \frac{\lambda_{res}}{n_g} \left( \frac{\partial n_{eff}}{\partial n_{core}} \frac{\partial n_{core}}{\partial \sigma} + \frac{\partial n_{eff}}{\partial n_{cladding}^{top}} \frac{\partial n_{cladding}^{top}}{\partial \sigma} + \frac{\partial n_{eff}}{\partial n_{cladding}^{bottom}} \frac{\partial n_{cladding}^{bottom}}{\partial \sigma} \right) \tag{6}$$

For different waveguide materials, different terms dominate the intrinsic sensitivity. The interaction coefficients, $\frac{\partial n_{eff}}{\partial n_X}$, can be determined with FEM simulations once the refractive index of the waveguide materials and the core geometry is known, where $n_X$ refers to different material domains of the waveguide.

We consider two different cases. Case 1: core materials with low photoelasticity, for example, Silicon (Si), Silicon Nitride (SiN), and Amorphous Silicon Carbide (a-SiC) waveguide cores with a highly photoelastic top cladding material such as PDMS, where the top cladding dominates $S_{int}$.

In this scenario, the effective index change is primarily influenced by the top cladding:

$$S_{int} = \frac{\lambda_{res}}{n_g} \left( \frac{\partial n_{eff}}{\partial n_{cladding}^{top}} \frac{\partial n_{cladding}^{top}}{\partial \sigma} \right) \tag{7}$$

Thus, the total sensitivity simplifies to:

$$S_{tot} = \psi P_{out} \frac{Q}{n_g} \left( \frac{\partial n_{eff}}{\partial n_{cladding}^{top}} \frac{\partial n_{cladding}^{top}}{\partial \sigma} \right) \tag{8}$$

Case 2: core materials with high photoelasticity, such as chalcogenide waveguide core, where core photoelasticity is comparable to that of PDMS top cladding photoelasticity.

In this case, since the waveguide mode is confined in the core material, only the core photoelasticity dominates the $S_{int}$ and cladding photoelasticity can be ignored:

$$S_{int} = \frac{\lambda_{res}}{n_g} \left( \frac{\partial n_{eff}}{\partial n_{core}} \frac{\partial n_{core}}{\partial \sigma} \right) \tag{9}$$

leading to the total sensitivity:

$$S_{tot} = \psi P_{out} \frac{Q}{n_g} \left( \frac{\partial n_{eff}}{\partial n_{core}} \frac{\partial n_{core}}{\partial \sigma} \right) \tag{10}$$

Eq.(10) and Eq.(8) show that $\psi$, $P_{out}$, $Q$, and $n_g$ are the scaling factors that affect the sensitivity for all kinds of waveguides in ultrasound sensing, and the core and cladding material photoelasticity determines the dominating factor of $S_{int}$ in ultrasound sensing.

We used a semi-analytical waveguide model [1] to calculate $n_{eff}$ of different waveguides, approximated Eq. 7 and Eq. 9 for different platforms and considered PDMS top cladding for each waveguide cross-section. The results of these calculations are given in Table 1. We observe that if the sensing is dominated by the photoelasticity of the cladding material, $S_{int}$ increases as the optical confinement of the mode decreases. Since the minimum bending radius of the waveguide increases as the optical confinement decreases, the result suggests that there is a trade-off between $S_{int}$ and the miniaturization potential of ring resonators. On the other hand, if the core material's photoelasticity dominates the sensitivity, then the overall sensitivity is higher than the waveguides, where the cladding photoelasticity is the dominating factor.

|  | Waveguide core material | | | | |
| --- | --- | --- | --- | --- | --- |
|  | Si | a-SiC | SiN | ChG | Unit |
| Core dimensions | $450 \times 220$ | $800 \times 280$ | $1000 \times 400$ | $2400 \times 850$ | nm×nm |
| $n_{core}$ | 3.476 [2] | 2.55 [3] | 1.996 [2] | 2.33 [4] | RIU |
| $n_{eff}$ | 2.3407 | 1.9030 | 1.608 | 2.203 | RIU |
| $n_g$ | 4.262 | 2.789 | 2.096 | 2.408 | RIU |
| $\frac{\partial n_{eff}}{\partial \lambda} \times 10^3$ | -1.24 | -0.572 | -0.315 | -0.132 | RIU/nm |
| $\frac{\partial n_{eff}}{\partial n_{core}}$ | 1.136 | 0.961 | 0.827 | 1.018 | RIU/RIU |
| $\frac{\partial n_{eff}}{\partial n_{cladding}^{top}}$ | 0.17 | 0.211 | 0.272 | 0.0234 | RIU/RIU |
| $\frac{\partial n_{core}}{\partial P}$ | n.a. | n.a. | n.a. | 545* [5] | RIU/TPa |
| $\frac{\partial n_{cladding}^{top}}{\partial P}$ | 664 [6] | 664 [6] | 664 [6] | 664 [6] | RIU/TPa |
| $\frac{\lambda_{res}}{n_g}$ | 364 | 556 | 740 | 644 | nm/RIU |
| $S_{int}$ | 40 | 78 | 134 | 410 | fm/kPa |

**Table 1** Theoretical calculation of intrinsic sensitivity for different waveguide cores with a highly elasto-optic polymer cladding such as PDMS. All calculations are made based on the TE0-like waveguide mode and at 1550 nm.
*Calculated using the values given for $C_R$ and $C_\theta$ in [5]
n.a. : not applicable since Eq. 7 is used for the $S_{int}$ calculation.
RIU : refractive index unit

# 3 k-Wave Simulations of Ultrasound Imaging Experiments

In this section, using the k-wave [7] add-on in MATLAB, we explain how the imaging in our setup is performed. The simulation and experiment results agree well regarding the time of flight between the initial pulse arrival and the signal from the samples.

In these simulations, $dx = dy =3.85\,\mu$m and $dt =0.125$ ns is defined as the k-wave grid parameters. Speed of sound is used as $1540\,\mathrm{m\,s^{-1}}$. A longitudinal wave equivalent pulse is defined at an angle representing the experiments. The speed of sound and the density of the single aluminum wire or the double fibers are defined at the approximate positions, which are found as a result of the imaging experiments (main text Fig. 5g, 5h). For both of the simulations, the grid parameters were kept the same; the simulation time window is adjusted according to the experiments.

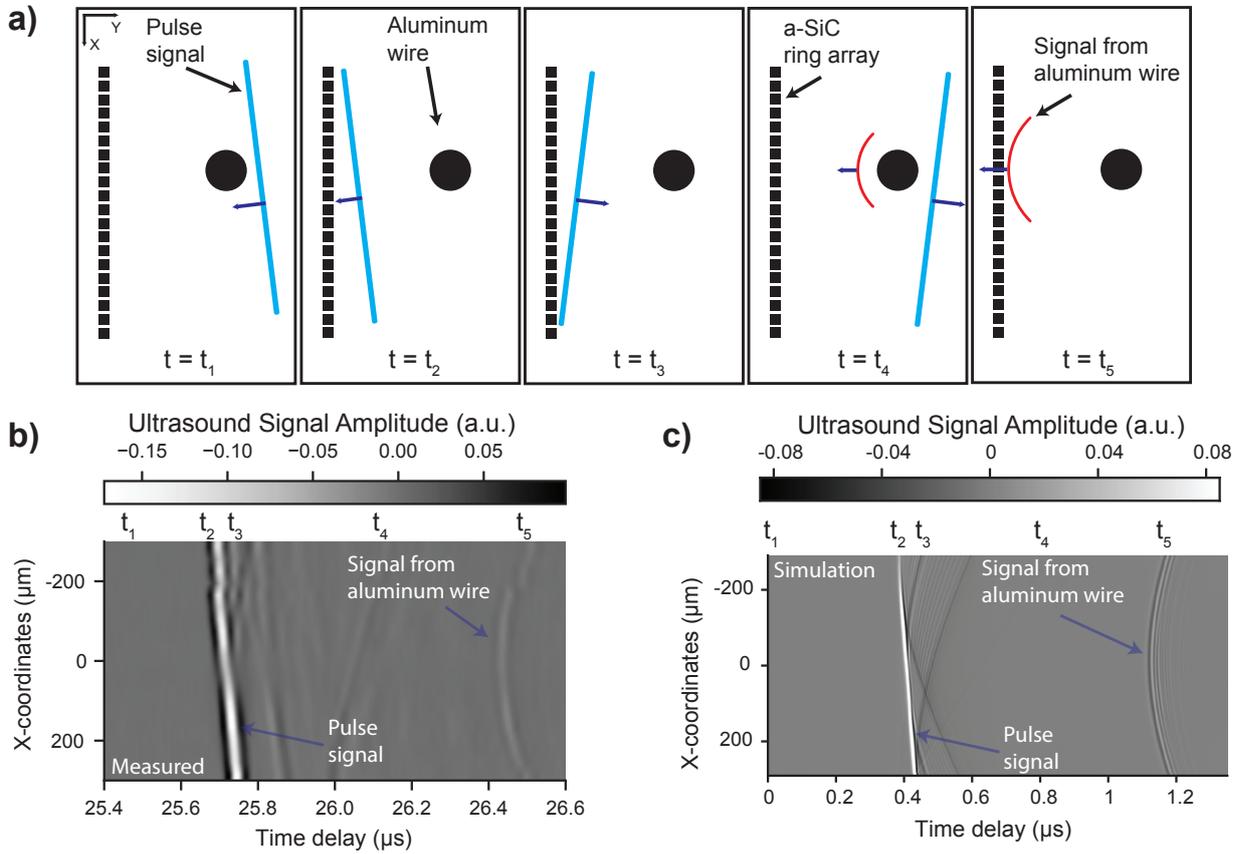

**Fig. S2** Comparison of the imaging experiments and k-wave simulations for the aluminum wire imaging experiment. **a)** schematic demonstration of the wave propagation in the water tank for different timestamps. **b)** Measured data using the linear array of 20 a-SiC ring resonators. **c)** Recorded amplitude data in the simulation by the point detectors.

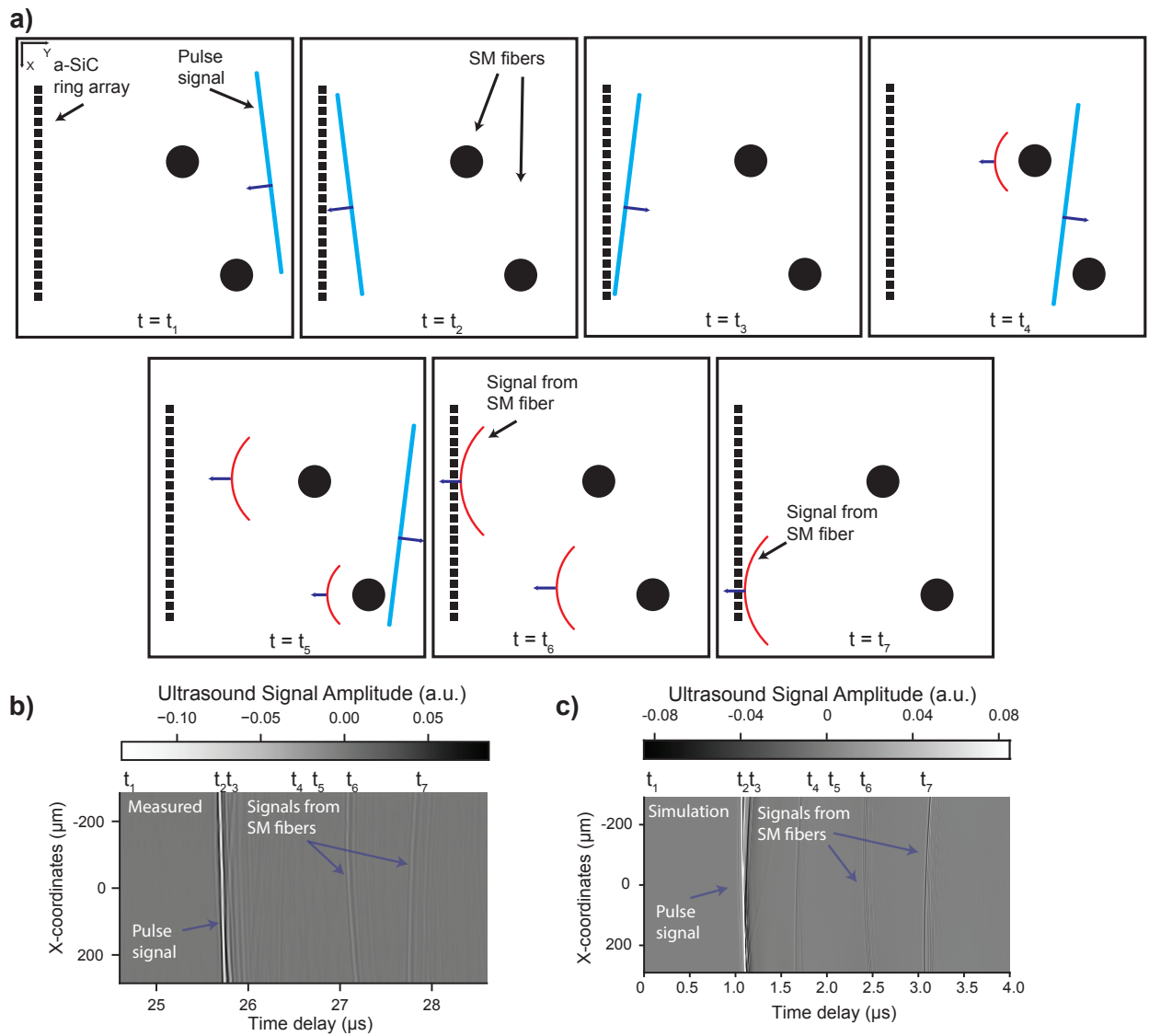

**Fig. S3** Comparison of the imaging experiments and k-wave simulations for the two SM fiber imaging experiments. **a)** schematic demonstration of the wave propagation in the water tank for different timestamps. **b)** Measured data using the linear array of 20 a-SiC ring resonators. **c)** Recorded amplitude data in the simulation by the point detectors.

8

# 4 Pressure calibration using a hydrophone

In this section, we expand on the details of the calibration of the pressures in our setup and the calculation method of the NEPD of the sensors. A calibrated needle hydrophone (0.075 mm, Needle Hydrophone, Precision Acoustics, SN:1302) is used to calibrate the pressure generated in our setup through the use of the immersion ultrasound transducer (V316-SM, Evident). The sensor and the hydrophone were placed at the same depth in both the calibration and NEPD characterization experiments. The in-plane position of the hydrophone of the sensor is optimized for the maximum amplitude received by the sensor or the hydrophone. A Gaussian signal is generated through the use of an AWG (Fig.S4a), and the same pulse is used in both experiments. The recorded signal using the hydrophone is plotted in Fig. S4b, and the FFT of this signal is plotted in Fig. S4c and Fig. S4d. FFT amplitude is converted to dB scale in Fig. S4d to visualize the calibration frequency high and low cut-off values. The calibration bandwidth is limited to the frequency cut-off values where the pressure signal is found to be 10dB less than the highest pressure signal generated by the ultrasound transducer for the lower cut-off, and the highest frequency of the calibration data of the hydrophone is used as the upper cut-off. Sensitivity data (Fig. S4e) of the hydrophone from the calibration certificate is used to calculate the pressure spectrum at the location of the hydrophone. Since the calibration data is sampled at 1 MHz, we fit a polynomial to this data to smooth the results. The calculated pressure spectrum is plotted in Fig. S4f and further used in the a-SiC ultrasound sensors' NEPD calculations.

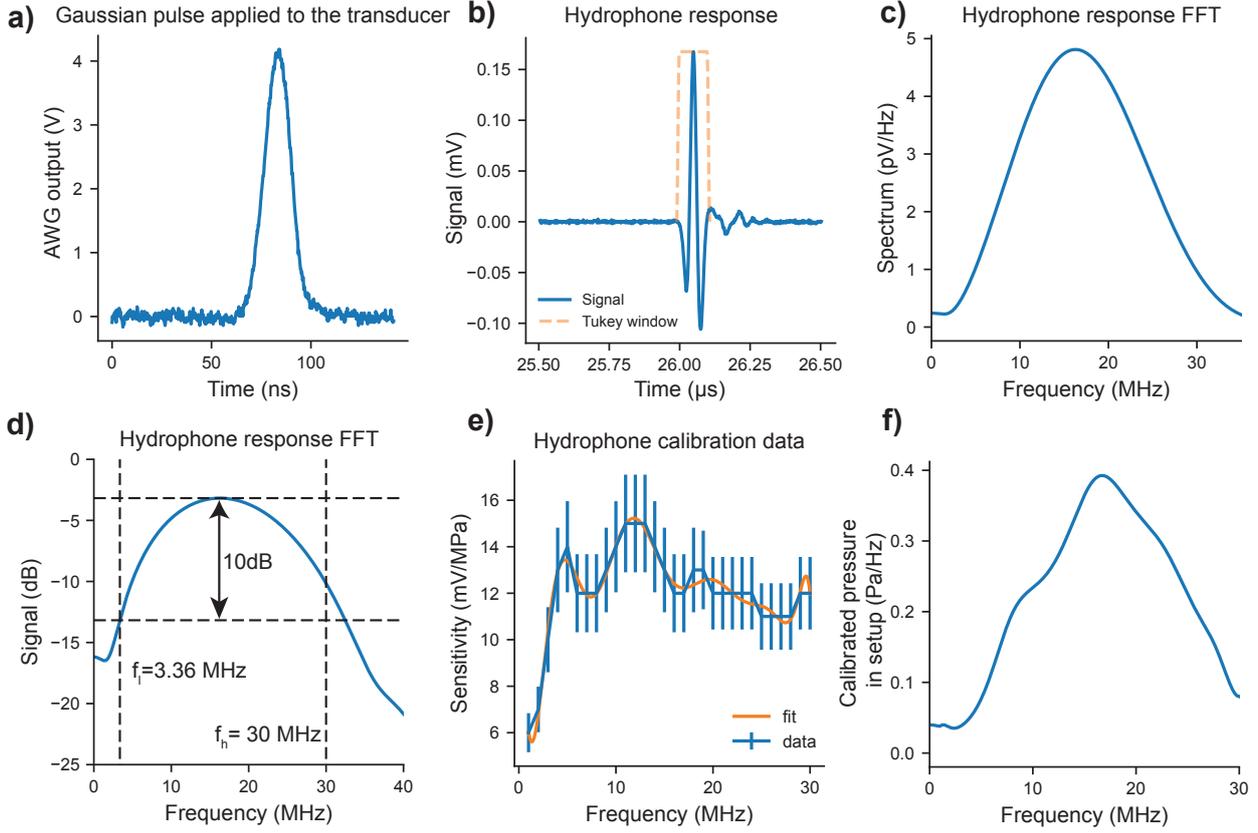

**Fig. S4** Pressure calibration of the ultrasound transducer. **a)** The Gaussian pulse applied to the ultrasound transducer for pressure calibration of the transducer. Measured from the output of the arbitrary waveform generator (AWG). **b)** The measured signal from the hydrophone, which is placed across the ultrasound transducer at a distance greater than the near field of the ultrasound transducer. **c)** FFT of the hydrophone signal shown in **(b)**. **d)** The calibration bandwidth determination from the FFT result of the hydrophone. **e)** The calibration data of the hydrophone, supplied by Precision Acoustics ltd. A fit is performed to interpolate the data points for further calculations of the NEPD and pressures. **f)** The pressure spectrum in the setup calculated using **(c)** and **(e)**.

## 5 NEPD calculation of the a-SiC ultrasound sensor

To characterize the a-SiC sensors NEPD in the bandwidth of the calibration we used the same Gaussian signal described in Fig.S4a, placed the sensor at the same depth as the hydrophone, and recorded the ultrasound response of the sensor. The response to the Gaussian pulse is plotted in Fig.5a and its FFT is given in Fig.S4b. Using the frequency response of the sensor and the calibrated pressure spectrum from Fig.S4f responsivity of the a-SiC ring resonator is calculated and the result is plotted in Fig.5c. Following the recording of the ultrasound signal with the sensor, the noise is recorded at the same wavelength and laser power as the ultrasound recordings. No averaging and no ultrasound signal is applied while recording the noise signal. This noise recording is plotted Fig.5d and its noise amplitude spectral density (ASD) (Fig.5e) of estimated using the Welch method with segments of 2048 samples, overlapping of 512 samples and using a Hanning window. By dividing the ASD (Fig.5e) to the responsivity (Fig.5c) of the sensor, the NEPD is calculated and plotted (Fig.5f). The maximum NEPD of the sensor in the range of calibration is found to be $55\,\text{mPa}/\sqrt{\text{Hz}}$ and the mean value is found to be $45\,\text{mPa}/\sqrt{\text{Hz}}$. The characteristic distribution of the hydrophone calibration data given in Fig.S4e
10

resembles itself in the NEPD distribution of the senor. This observation suggests that the sensor's response is approximately flat over the range of calibration frequencies.

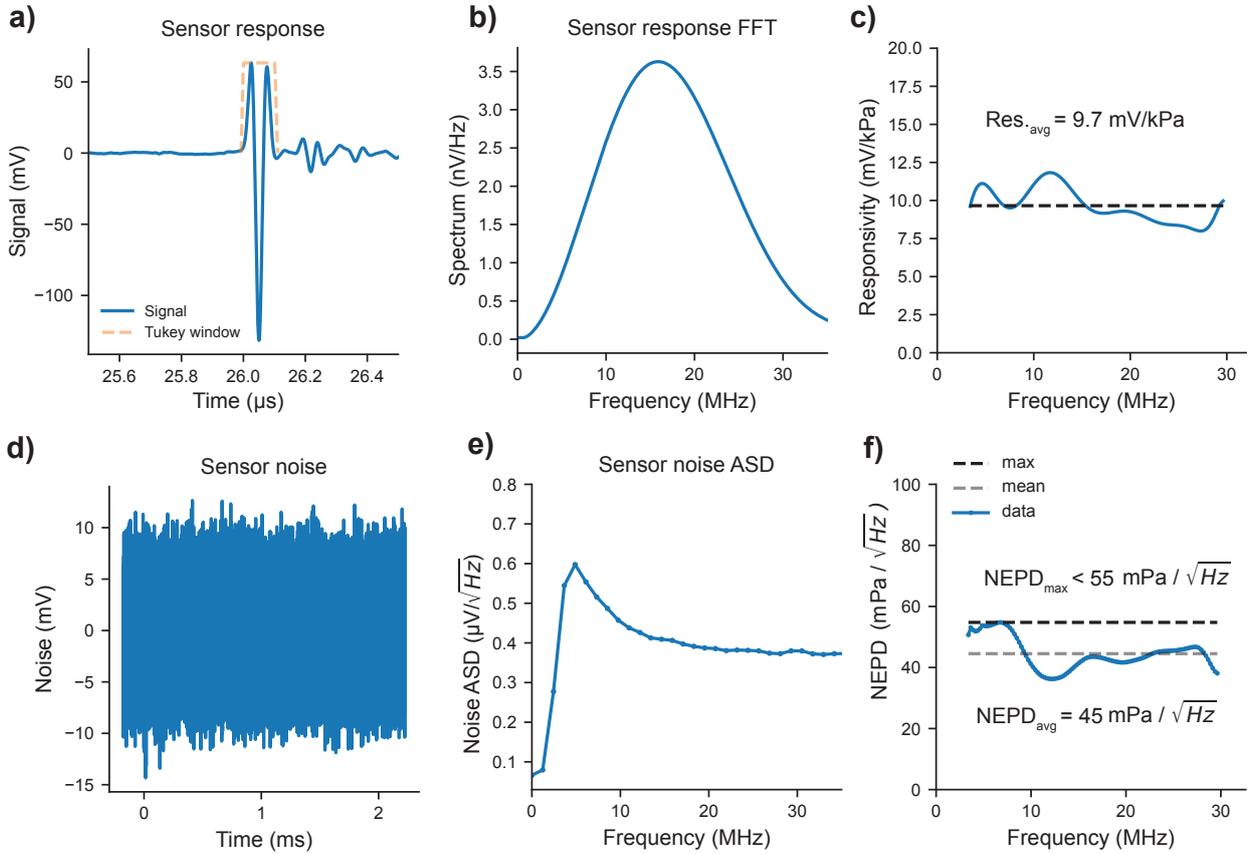

**Fig. S5** NEPD calculation of the a-SiC ring resonator. **a)** The measured signal from the a-SiC microring resonator represented in the main text Fig 3., which is placed across the ultrasound transducer at the same distance as the hydrophone. **b)** FFT of the signal shown in **(a)**. **c)** The calibrated responsivity function of the a-SiC microring resonator in ultrasound sensing. **d)** Recorded noise signal without averaging or any ultrasound pressure application. **e)** The noise amplitude spectral density (ASD) of the signal shown in **(d)**. The Welch method is used to calculate the ASD with a Hanning window with segments of 2,048 samples, 512 samples overlapping, and a linear detrend. **f)** The noise equivalent pressure density of the a-SiC microring resonator is calculated using the responsivity spectrum and noise amplitude spectral density calculation. The mean and maximum value of the NEPD are also annotated with dashed lines and text.

# 6 Determination of the lateral position of the ring resonators on the chip

In the main text, we annotate the ultrasound measurement range of wavelengths with the array of ring resonators in Fig. 4b. The ultrasound data with each ring resonator in the array is collected from the indicated 20 notches, sequentially. However, the order of the resonance frequencies differs from the placement order of the ring resonators in the chip. This can happen due to the variations in the fabrication during the EBL exposure process, unequal etching, or non-uniformity in the cladding PDMS layer. We correct the order of recorded ultrasound signals and determine the position of the ring resonators by using the two ultrasound experiments as depicted in Fig.S6a, a recording of the ultrasound pulse without a sample in the water, and Fig.S6b, the recording of the ultrasound pulse with the aluminum wire between the ultrasound transducer and the ring array. Fig.S6c and Fig.S6d are the recorded ultrasound signals from the ring resonator array, where the resonator index is given as the measurement order (i.e., resonance wavelength order). Both of the recordings show that, according to the initial pulse arrival time, the ring resonators from index 16 to 19 receive the pulse earlier than expected, suggesting that their placement order in the chip does not follow their resonance wavelength; hence, their index needs to be corrected. Accordingly, we sort the indices of recorded responses given in Fig.S6c and Fig.S6d based on the initial pulse arrival time in Fig.S6e and Fig.S6f, respectively. Furthermore, we realize that the initial pulse arrival sorting is not enough to correctly sort the indices of the two resonators, namely the resonator at indices 4 and 5, in Fig.S6e and Fig.S6f. As indicated in the figures with the red arrow, the time of receiving the reflection signals of the initial ultrasound pulses for these resonators creates a discontinuity in the signal recorded, which suggests a surface defect near the location of the resonators that is causing the signal to be delayed further than uniform over the whole array. According to this observation, we switch the indices of these two resonators and reach the final, correctly sorted indices of the whole array. Fig.S6g and Fig.S6h plot the resulting final data from both of the experiments depicted in Fig.S6a and Fig.S6b, respectively.

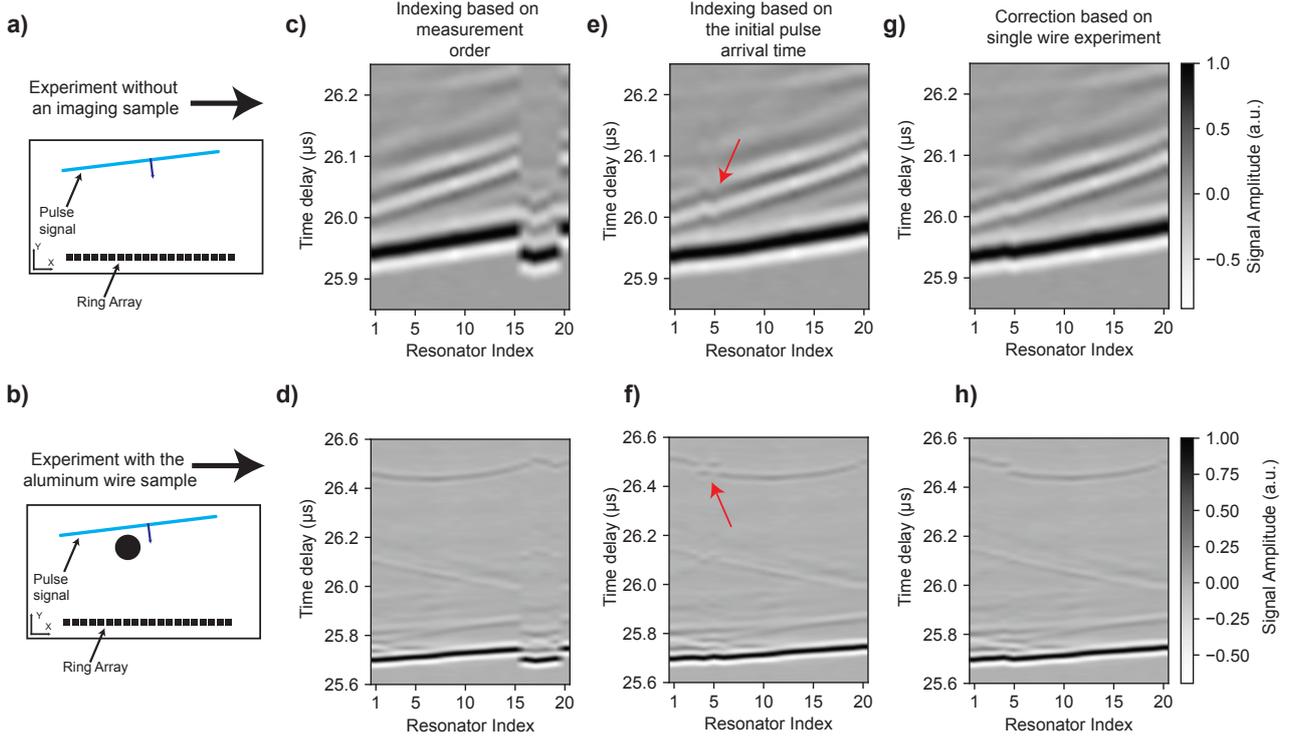

**Fig. S6** Sorting of the array of ring resonators using the experimental data with and without an aluminum wire to determine the location of the ring resonators on the chip. **a**) Schematic cross-sectional view of the pulse sensing experiment with the ring resonator array. **b**) Schematic cross-sectional view of the imaging of the aluminum wire experiment with the ring resonator array. **c**) and **d**) The recorded ultrasound signals, which are sorted based on the notch wavelength order in the time domain for the experiments shown in (**a**) and (**b**), respectively. **e**) and **f**) The time traces in (**c**) and (**d**) are sorted based on the initial pulse arrival time. The discrepancy in the reflection signals is indicated with red arrows. **g**) and **h**) The time traces after the final adjustment based on the reflection signals in (**e**) and (**f**)

\AtBeginShipout\AtBeginShipoutDiscard

\end{document}